\documentclass[vecphys]{svmult}

\usepackage{makeidx}         
\usepackage{graphicx}        
\usepackage{epsfig}                             
\usepackage{multicol}        
\usepackage[bottom]{footmisc}
\makeindex             

\begin{document}
\title*{ Mining the Local Volume}
\titlerunning{Mining the Local Volume}

\author{Igor D. Karachentsev\inst{1}
\and Valentina Karachentseva\inst{2}
\and Walter Huchtmeier\inst{3}
\and Dmitry Makarov\inst{1}
\and Serafim Kaisin\inst{1}
\and Margarita Sharina\inst{1}
\and Lidia Makarova\inst{1}}
\authorrunning{Igor D. Karachentsev et al.}
\institute{Special Astrophysical Observatory, Russian Academy of Sciences,
 N. Arkhyz, 369167,  Russia
\texttt{ikar@sao.ru,dim@sao.ru,skai@sao.ru,sme@sao.ru,lidia@sao.ru}
\and Astronomical Observatory, Kiev National University, Kiev, 04053 Ukraine                                     \\
\texttt{vkarach@observ.univ.kiev.ua}
\and Max-Planck-Institut f\"{u}r Radioastronomie, Auf dem H\"{u}gel 69, D-53121 Bonn, Germany
\texttt{p083huc@mpifr-bonn.de}}

\maketitle

\begin{center}
{\bf Absrtact.}
\end{center}
 {\footnotesize \em After recent systematic optical, IR, and HI surveys, the total number of
known galaxies within 10 Mpc has increased from 179 to 550. About half this
Local Volume (LV) sample is now been imaged with HST, yielding the galaxy
distances with an accuracy of about 8\%. For the majority of the LV galaxies
we currently have $H\alpha$ fluxes that allow us to reconstruct the star
formation history of our neighbourhood.

   For the late-type LV galaxies their HI masses and angular momentum
follow the linear relation in the range of 4 orders, which is expected for
rotating gaseous disks being near the gravitational instability threshold.

   The data obtained on the LV galaxies imply important cosmological
parameters, in particular, the mean local matter density and HI mass
density, as well as SFR density.

  Surprisingly, the local Hubble flow around the LV groups is very quiet,
with 1D rms deviations of 25 km s$^{-1}$,which is a signature of the
Universe vacuum-dominated on small scales. The cold infall pattern around
nearby groups provides us with a new method to determine the total mass of
the groups independent from virial mass estimates. }

\section{The Local volume census.}

The first step towards compiling a Local volume (LV) sample of galaxies
situated within 10 Mpc was made by Kraan-Korteweg \& Tammann [1]
who published a list of 179 nearby galaxies with radial velocities
$V_{LG} < 500$ km s$^{-1}$. Then Karachentsev [2] updated their
list to 226 objects. Later, fast increasing the LV sample was proceeding
by different ways:
 a) via Z-surveys of the known ( UGC, MCG, CGCG ) catalogues;
 b) based on special searches for dwarf members of nearby groups
    around M~31, M~81, IC~342, Cen~A, NGC~253, M~94, M~101, NGC~6946;
 c) from searches for LSB galaxies in wide sky regions;
 d) via special HI and NIR surveys in the Zone of Avoidance;
 e) from blind HI sky surveys (Staveley-Smith et al.[3], Koribalski et al. [4]).

  During the last decade, a comprehensive study of the LV galaxies
was undertaken by Karachentseva \& Karachentsev and their co-workers.
The basic steps of this long-term project are listed in Table 1.

\begin{table}
\centering
\caption{Basic stages of the project}
\begin{tabular}{p{4cm}p{2.5cm}p{1cm}l} \\ \hline
       Item              &    Means           & Status&       Results                 \\
\hline
  All-sky search for new &  POSS-II,          & 100\% &  350 new LSB dwarf              \\
  Local volume members.  &  ESO/SERC.         &       & galaxies were found.            \\
  HI line survey of 600  &  Effelsberg,       & 100\%  & 100 new LV dwarfs              \\
  dwarfs from KK-lists.  &  Nancey, ATCA.     &       & with $V_{LG} <$ 550km s$^{-1}$. \\
  CCD (B,V,R)- imaging   &  6m SAO,           &  50\%  & 150 LV members             \\
  of the LV galaxies.   &  2.5m Nordic.      &       & resolved into stars             \\
			 &                    &       & for the first time.              \\
  Distance measurements  &  HST.              &  50\%  & accurate TRGB distances       \\
  to the LV galaxies.    &                    &       & to 200 galaxies.                \\
  $H\alpha$ imaging all  &  6m SAO,           &  60\%  & HII-pattern and SFR          \\
  the LV galaxies.       &  2.2m ESO.         &       & for 300 galaxies.               \\
  HI velocity field for  &  GMRT              &  80\%  & Dark Matter properties         \\
  90 tiny dIrrs [5].     &                    &       & on scales of $\sim$1 kpc.            \\
\hline
\end{tabular}
\end{table}

General observational data on 451 galaxies were summarized in the Catalog
of Neighboring Galaxies by Karachentsev et al. [6]. The present version
of the LV sample contains 550 objects.

  As it is known, the simpler selection criterion taken for any sample,
the easier interpretation of the sample properties. This Heisenberg's
principle of (un)certainty gives a great advantage to the LV sample because
of its simplest selection criterion, $D < 10$ Mpc.

  However, an improving census of the Local Volume has a two-way traffic.
Modern automated redshift surveys, like SDSS, 2dF, DEEP2, 6dF, produce
a lot of spurious objects with radial velocities around zero.
For instance, the DEEP2 survey generated about 700 false members of the LV
fixed in LEDA and NED. There are also some cases in these databases, like
"dIrr galaxy" AM 0912-241 with $V_h = +614$ km s$^{-1}$ which turns out to be
a photographic emulsion defect.

\section{Mapping galaxy distribution and peculiar velocities in the LV.}

  The sky distribution of the LV galaxies looks extremely inhomogeneous
owing to the presence of galaxy groups [7] and voids
[8]. Spatial distribution of the nearest
galaxies inside and around the Local group (LG) is presented in Fig. 1
in the Supergalactic coordinates. Galaxies of different linear diameters are
shown by different size balls. The circle indicates the radius of zero-
velocity surface ($R_o = 0.9$ Mpc) separating the LG against the cosmic
Hubble expansion. The edge-on view of the LG and its suburbs (lower panel)
demonstrates larger scatter of dwarf galaxies with respect to the SG plane.

\begin{figure}
\centering
\includegraphics[height=10cm]{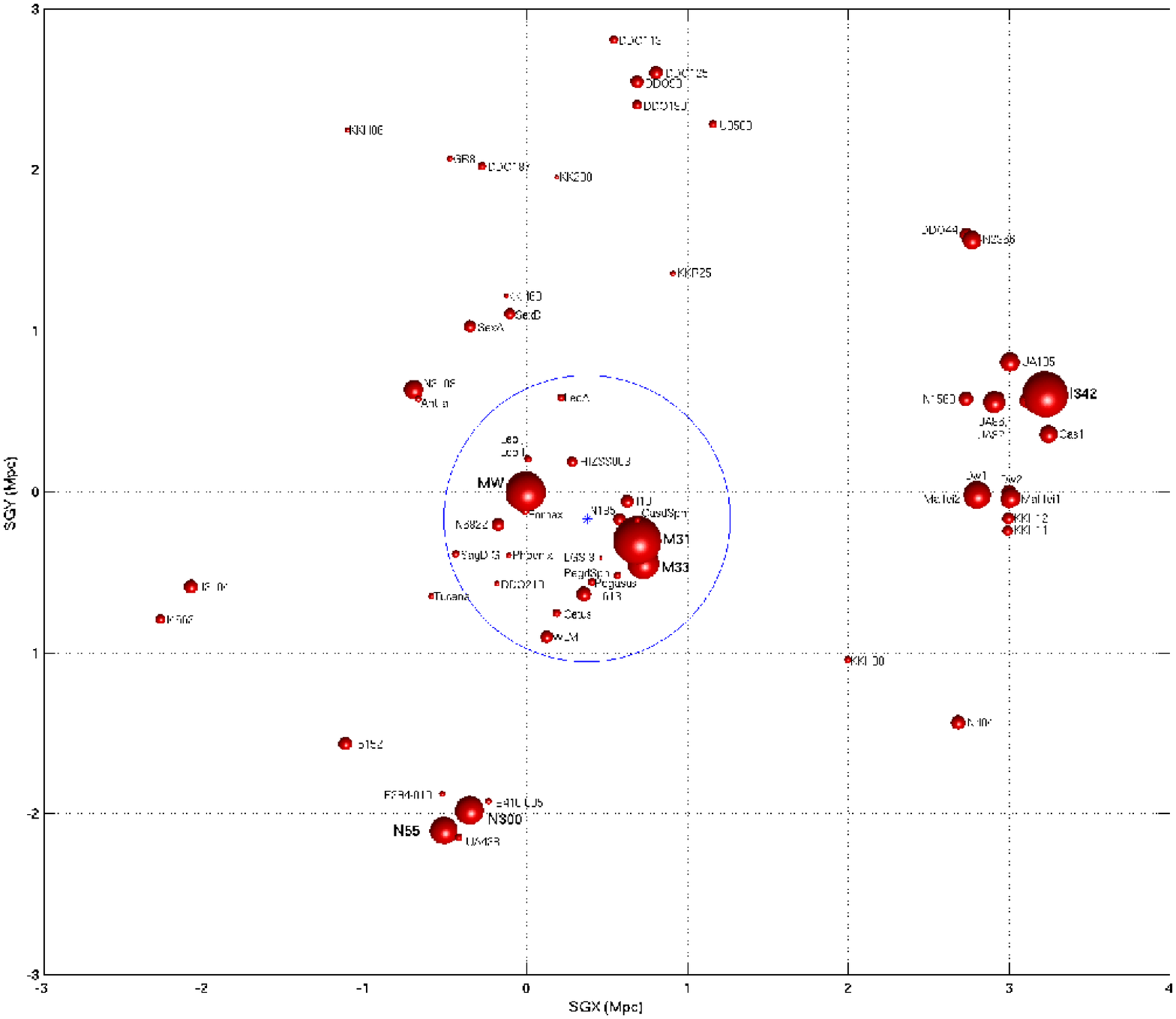}
\includegraphics[height=10cm]{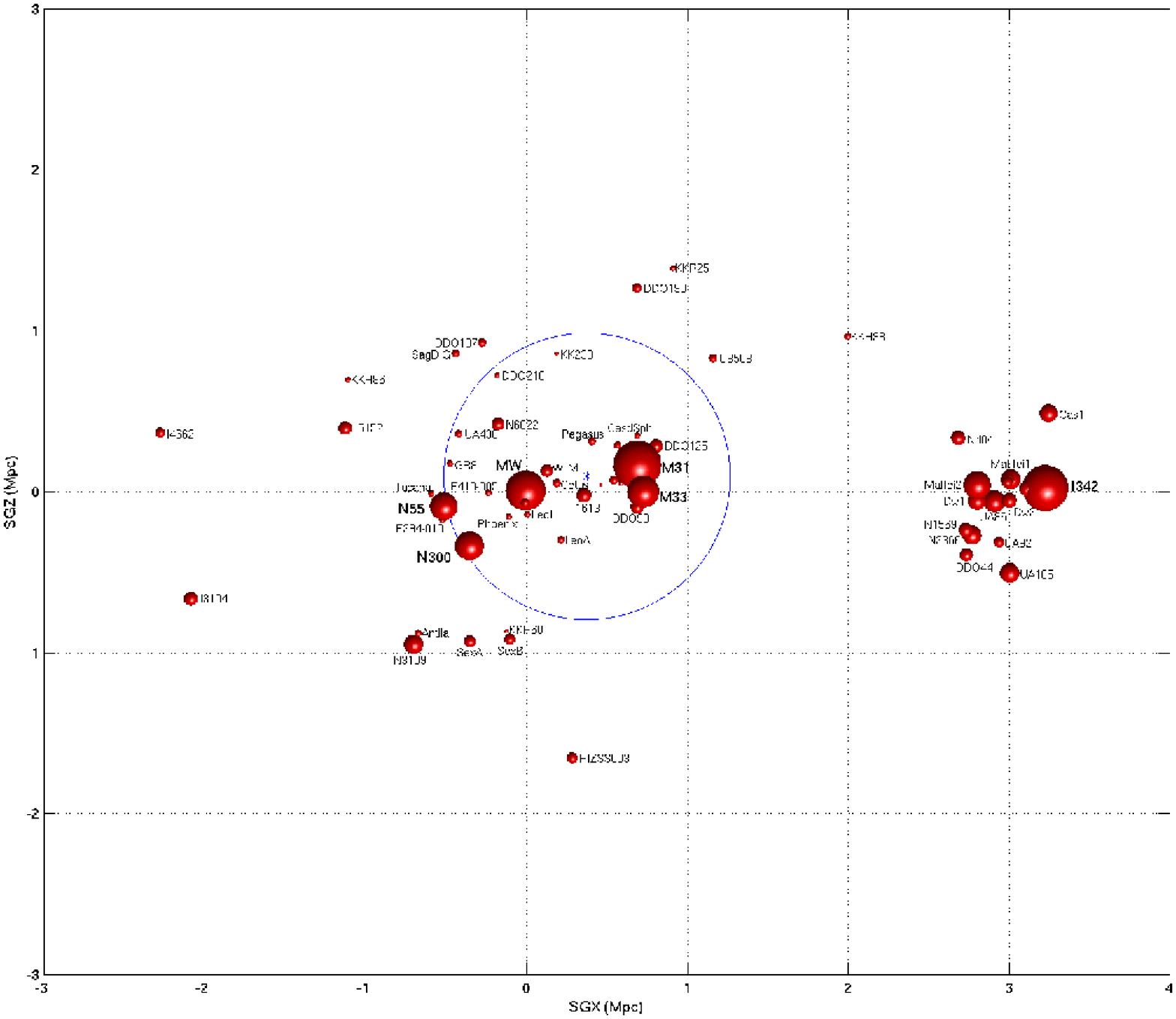}
\caption{Spatial distribution of the nearest galaxies
inside and around the Local group in the Supergalactic coordinates}
\end{figure}

  While having accurate velocities and distances for ~200 LV galaxies, one
can study distribution of peculiar velocities within the LV. Fig. 2
presents a peculiar velocity map for the LV galaxies in the equatorial
coordinates given in the LG reference frame and smoothed with a window of
15$^\circ$. The distribution shows the local Hubble flow to be generally
quiet ($\pm$30 km s$^{-1}$) with a small area of negative peculiar velocities
~ --250 km s$^{-1}$  in the direction towards the Leo
constellation. This phenomena can be caused by
the apparent motion of the Local Sheet as a whole from the large Local
void towards the neighboring Leo cloud [9]. Note that the
local peculiar velocity field observed on a scale of 10 Mpc has not any
relation to the so-called Virgo-centric infall often used by different
authors to "improve" distances to nearby galaxies via their velocities.
\begin{figure}[bth]
\centering
\includegraphics[height=6cm]{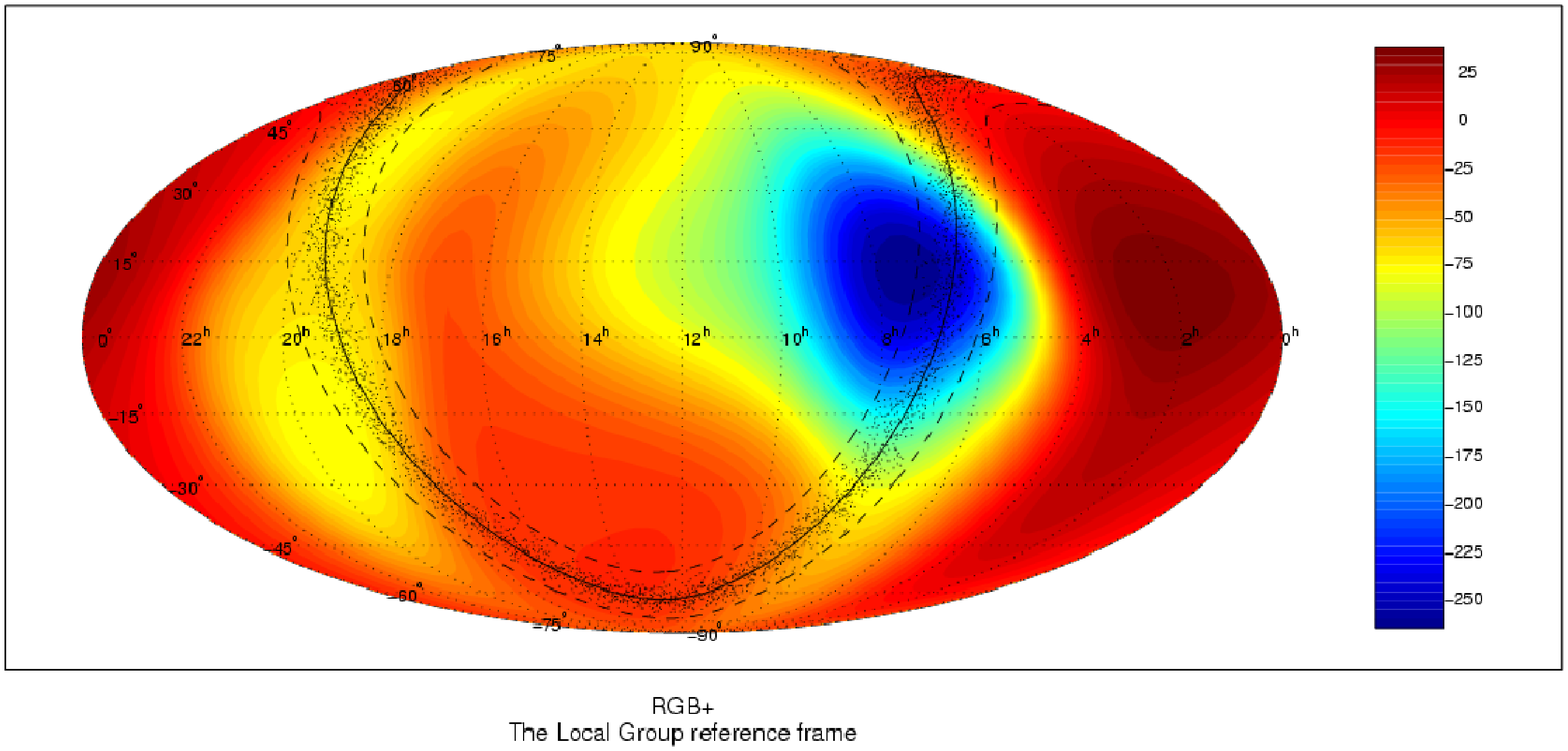}
\caption{Peculiar velocity map for the LV galaxies}
\end{figure}

\section{Some basic relations for the LV galaxies.}

 Besides the global (external) Hubble law, galaxies follow another
internal "Hubble law-2", the linear relation between their rotation
velocity and standard radius:

   $   V = h\cdot R, $ \\
where  $h = 137H_0$, or $1/h = 100$ Myr. This leads to the known empirical
relations between total mass $M_t$, luminosity $L$, total angular momentum
$J$, and surface brightness SB of a galaxy:

$    M_t\propto L \propto R\cdot V^2 \propto R^3,$  \\
meaning that the average spatial densities of giant, normal, and dwarf
galaxies are almost the same;

$    L \propto~ V^3, $                                \\
corresponding on a logarithmic scale to the Tully-Fisher relation;

$    J \propto M\cdot V\cdot R \propto M^{5/3},   $ \\
which is known as Muradyan law valid for a much wider range of cellestial
bodies from asteroids to galaxy superclusters [10];

$    SB \propto L/R^2 \propto R, $                    \\
Binggeli-Grebel relation (valid also for E and dSph galaxies), which
makes one search for extremely faint
dwarf galaxies among the objects of the lowest
surface brightness. These empirical scaling relations probably mean that
stellar population of dwarf and giant galaxies reside in dark matter
"corsets" of a standard profile.

  As it was shown by Roberts [11], hydrogen masses of disk-like
galaxies follow the empirical relation:

$     M_g \propto R^2, $                           \\
i.e. the mean HI surface brightness of giant and dwarf disks is almost
the same. This Roberts law leads to the following scaling relations:

$    M_g/M_t \propto R^2/R^3 \propto 1/R \propto M_t^{-1/3},$ \\
meaning that dwarf galaxies are relatively more gas-rich systems, being
capable of longer star formation activity than giant ones;

$    M_g \propto R\cdot V \propto j, $ \\
this Zasov relation between total hydrogen mass and specific angular
momentum of galaxies, shown in Fig. 3, means that galaxy disks are situated
in the state of equilibrium just above the threshold of gravitational
instability driving their star formation processes [12].
\begin{figure}
\centering
\includegraphics[height=10cm,angle=-90]{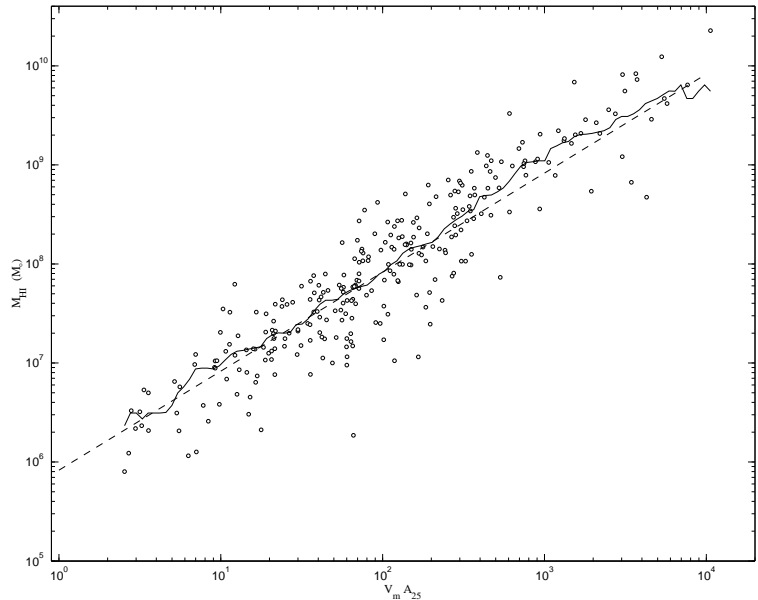}
\caption{The total hydrogen mass vs. specific angular momentum of galaxies}
\end{figure}

\section{$H\alpha$ flux and SFR for the LV galaxies.}

  Systematic $H\alpha$ imaging made for the LV galaxies allows us to measure
the star formation rate (SFR) for them in an unprecedented wide range. As
distinct from other samples, the LV sample offers an unique opportunity to
study the SFR for galaxies of different types and in different environment
without significant selection biases.

  Left and right panels of Fig. 4 present the SFR of nearby galaxies versus
their blue absolute magnitude and hydrogen mass, respectively. The dashed
lines correspond to a constant SFR per unit luminosity or unit hydrogen mass.
As can be seen from the left diagram, most of the galaxies brighter than
$-13^m$ follow a linear regression with the constant specific SFR. However,
SFRs and hydrogen masses demonstrate a steeper relationship [SFR]
$\propto{\cal M}^{3/2}_{HI}$, shown on the right panel by the solid line.
This feature seems to be in harmony with the following considerations
noted by Tutukov [13]. According to Schmidt law for local SF sites,
their rate of transformation of gas into stars is proportional to the
square of gas density:

$   d(n_g)/dt \propto (n_g)^2 .$ \\
Taking into account the above mentioned relations between the galaxy disk
parameters, we find that

$ SFR \propto d(n_g)/dt\cdot R^3 \propto (n_g)^2\cdot R^3 \propto (M_g)^2/R \propto M_g^{3/2}, $   \\
i.e. obtain the known Kennicutt law, but for the galaxies themselves,
not for individual HII regions only. Therefore, evolutionary history of
disks of galaxies looks to be driven mainly by SF processes.

  To characterize the past and the future evolution status of a galaxy,
we introduce two dimensionless parameters:

$p_*=\log {[SFR]\cdot T_0/L_B}\,\, {\rm and} \,\,f_*= \log {M_{HI}/[SFR]\cdot T_0}.$ \\

\begin{figure}
\centering
\includegraphics[height=6cm]{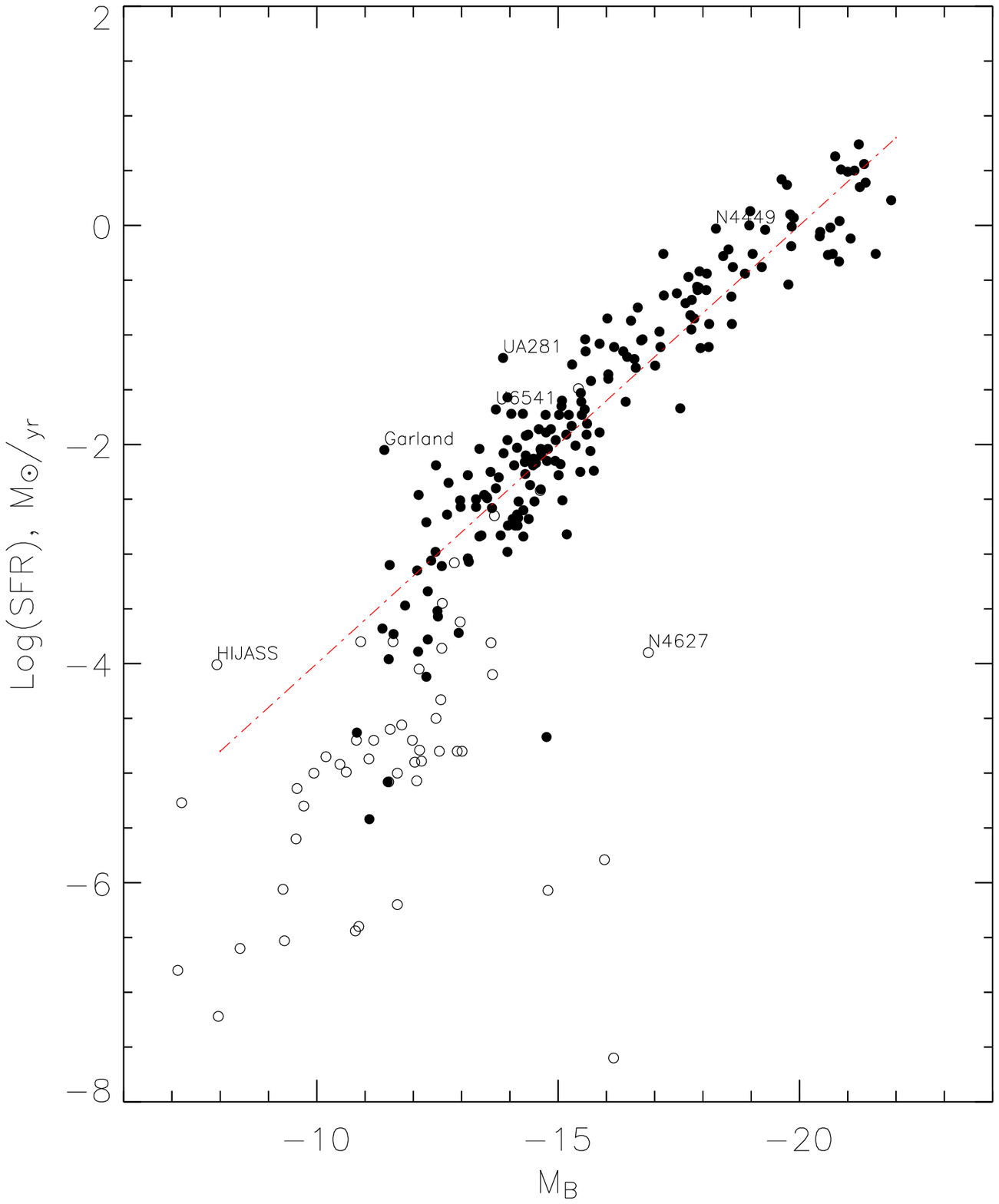}
\includegraphics[height=6cm]{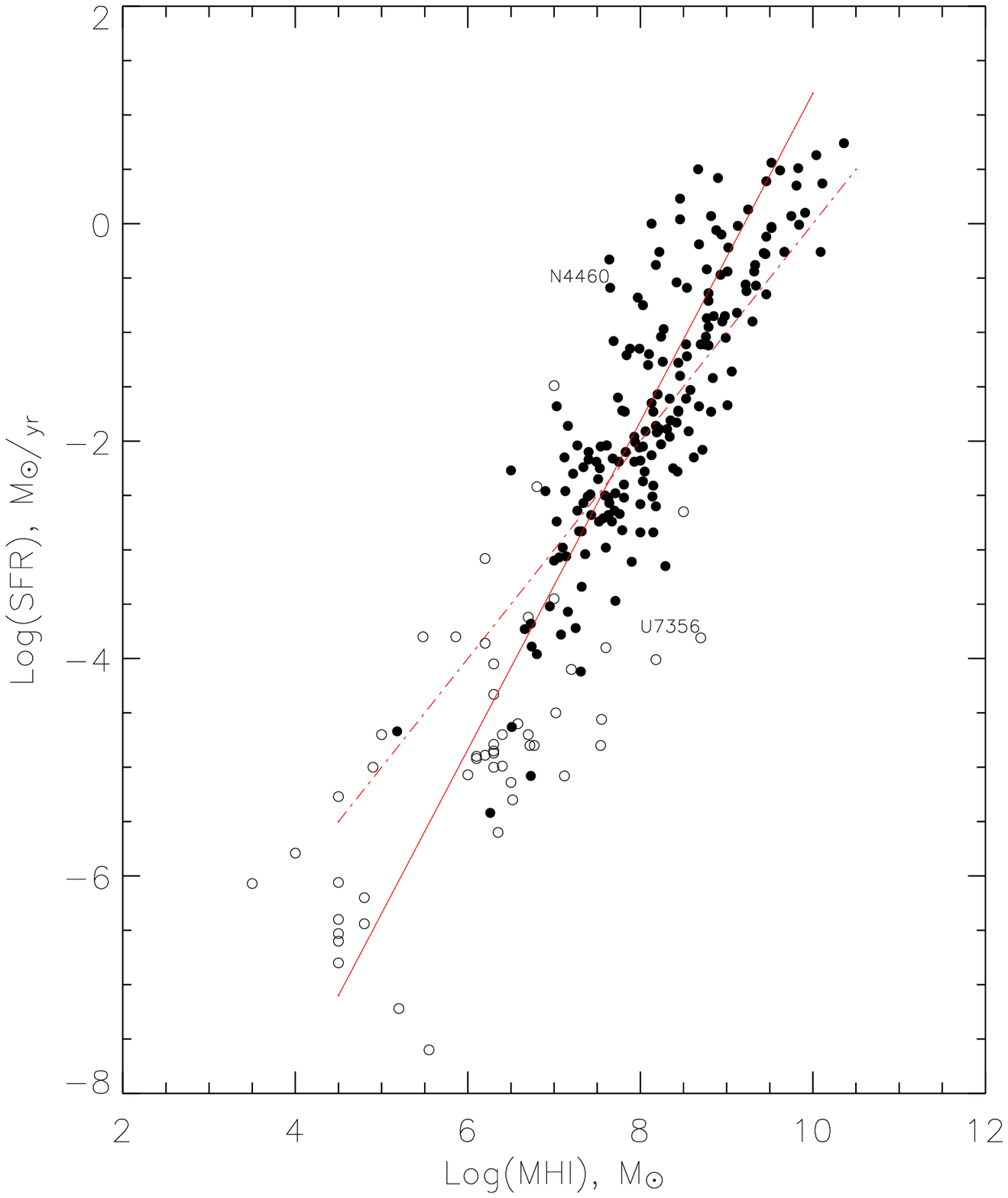}
\caption{The SFR of nearby galaxies vs. their blue absolute magnitude (left)
 and hydrogen mass (right)}
\end{figure}
The former parameter describes the galaxy formation timescale, the latter
shows the gas depletion timescale, both expressed in the Hubble time units,
$T_0$. The distribution of the LV galaxies on the ${p_*, f_*}$-plane is
displayed in Fig. 5. As a whole, the LV population concentrates around the
origin ($p_*=0,f_*=0$) of this diagnostic diagram. It means that the
\begin{figure}
\centering{\psfig{figure=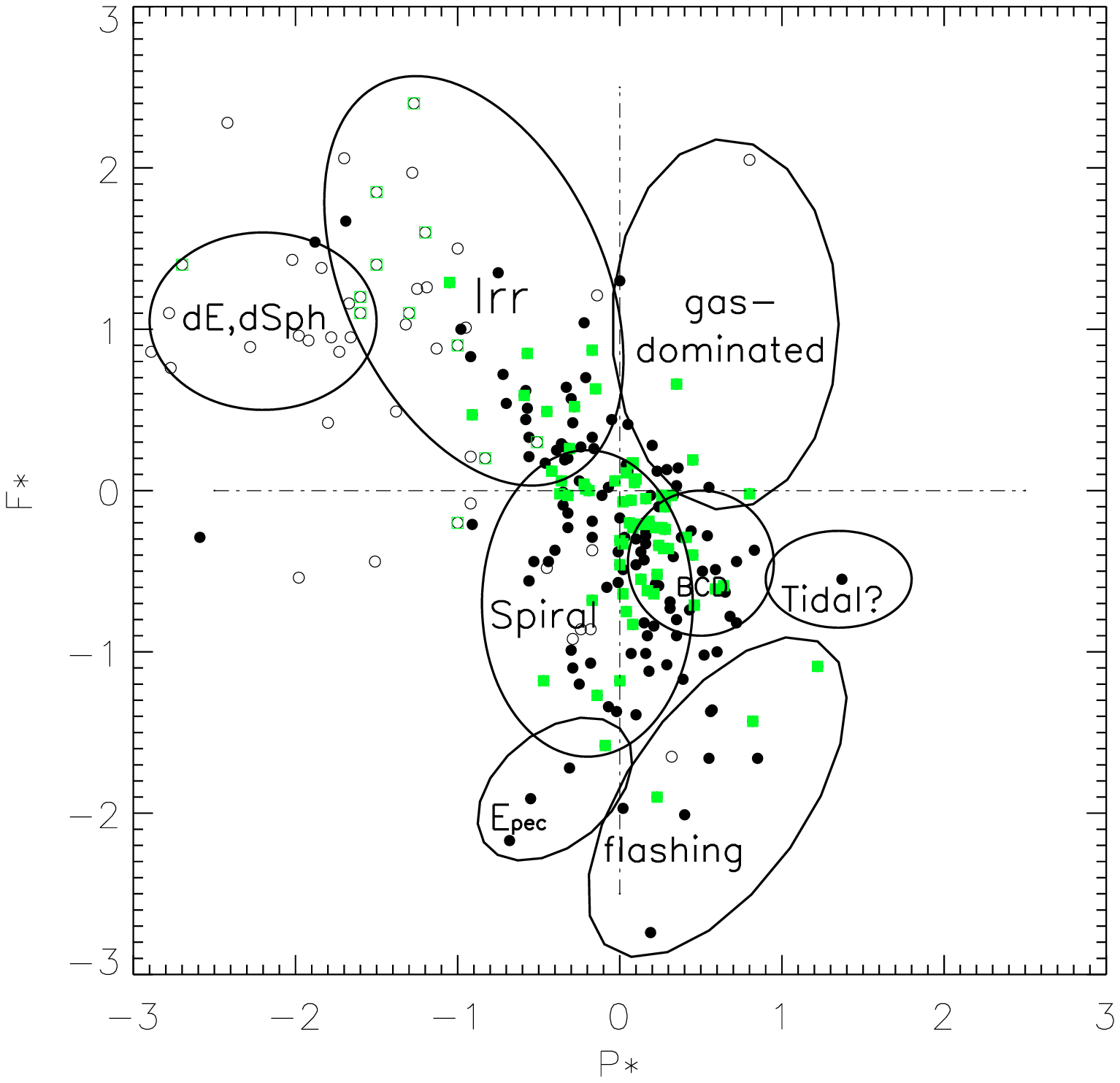,angle=0,width=10cm}}
\caption{Distribution of the LV galaxies on the ${p_*, f_*}$-plane}
\end{figure}

\begin{figure}
\centering{\psfig{figure=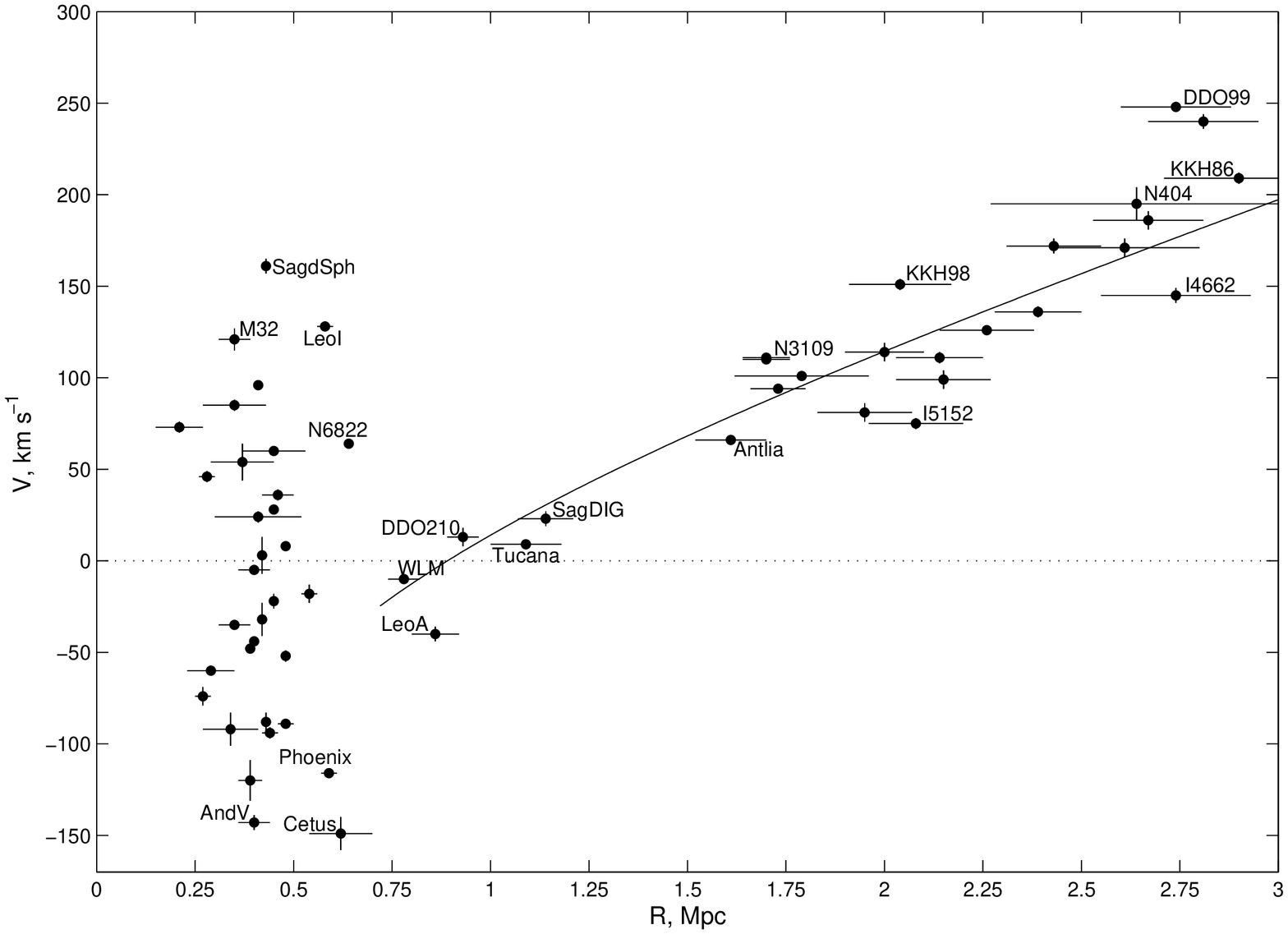,angle=-90,width=12cm}}
\caption{The local Hubble flow in the immediate surroundings of
the Local group}
\end{figure}

observed luminosity of a typical LV galaxy may be reproduced with its
observed SFR, and a typical LV galaxy has enough gas to continue
its observed SFR during the next Hubble time. However, galaxies of
different morphological types occupy different regions on the ${p_*,f_*}$-
diagram that must be a subject of elaborate analisys.

  It is generally accepted that the enhanced star formation in galaxies
is triggered by their interaction. But we did not find clear evidence for
such a suggestion. Curiously, the strongly disturbed tidal dwarf "Garland"
near NGC~3077 and the very isolated blue galaxy UGCA~281 have almost the
same extremely high specific SFR (see left pannel in Fig.4).

\section{Basic properties of the nearest groups.}
Over the last few years, searches for new nearby dwarf galaxies and
accurate distance measurements for them lead to a significant increase
of the population of known neighboring groups. Table 2 presents basic
parameters of 7 nearest groups situated within 5 Mpc from us: distance
to the group centroid, number of the group members with known radial
velocities, 1D velocity dispersion, projected radius, crossing time,
virial mass, total blue luminosity and virial mass-to-luminosity ratio.

\begin{table}
\centering
\caption{Basic parameters of the nearest groups}
\begin{tabular}{lccccccc} \\ \hline
 Parameter       &   M.Way  &  M31  &   M81  &  CenA  &  M83   & IC342  & Maffei   \\
\hline
 D,       Mpc    &   0.01   & 0.78  &  3.63  &  3.66  &  4.56  &  3.28  &  3.01    \\
 Nv              &    18    &  18   &   24   &   29   &   13   &   8    &   8      \\
$\sigma_v$,km/s  &     76   &   77  &    91  &   136  &    61  &    54  &    59    \\
$ R_p$,     Mpc    &   .16    & .25   &  .21   &  .29   &  .16   &  .32   &  .10     \\
 $T_{cross}$, Gyr    &   2.1    & 3.3   &  2.3   &  2.2   &  2.7   &  5.9   &  1.8     \\
 $M_{vir}$,    $10^{10}$ &   95     & 84    & 157    & 725    &  86    &  76    &  100     \\
 $L_B$,      $10^{10}$ &   3.3    & 6.8   &  6.1   &  6.0   &  2.5   &  3.2   &   2.7    \\
 $M/L$,     solar  &   29     & 12    &  26    & 121    &  34    &  24    &   37     \\
\hline
\end{tabular}
\end{table}
Remarkably, centroids of the groups have the radial velocity dispersion
around the local Hubble flow of only 25 km s$^{-1}$.

  Precise measurements of distances and radial velocities for galaxies
surrounding a group permits us to determine the radius of zero - velocity
surface, $R_o$, which separates the group from the global cosmic expansion,
and then the total mass of the group defined by Lynden-Bell [14] as

    $M_t = (\pi^2/8G)\cdot R_o^3\cdot {T}_o^{-2}.$ \\
Fig.6 exhibits the local Hubble flow in the immediate surroundings of
the Local group. Distances and velocities are given with respect to the
LG centroid. Similar "cold" velocity patterns are also seen around other
nearest groups, yielding $R_o$ values within 0.7 - 1.4 Mpc.

\section{Some LV parameters important for cosmology.}

  In spite from the presence of local voids, the average density of
luminosity within the radius of 8 Mpc around us exceeds 1.5 - 2.0 times
the global luminosity density [6]. Almost the same
excess is also seen in the local HI mass density [15].
About 2/3 of the LV galaxies belong to the known virialized groups like
the LG. Because the average virial mass-to-luminosity ratio for them is
40 $M_{\odot}/L_{\odot}$, the mean local mass density within 8 Mpc turns out to be
0.10 in units of the global critical density. This quantity is 2 - 3 times
as low as the global density of matter, $\Omega_m$ = 0.27. To remove
the discrepancy between the global and local quantities of $\Omega_m$,
we assume that the essential amount of dark matter (~70\%) exists
outside the virial radius of the groups.

  It should be stressed that the number density of test particles (dwarf
galaxies) in the LV is much higher than in any other distant volumes.
Therefore, systematic investigating the LV has great advantage in probing
the dark matter distribution on scales of 0.3 - 3 Mpc. In this respect
we note that the sum of virial mass for 7 nearest groups (around the
Milky Way, M 31, M81, CenA, M83, IC342, and Maffei) consists of
$1.3\cdot 10^{13} M_{\odot}$. But the sum of their total masses estimated via $R_o$
from external galaxy motions is $0.86\cdot 10^{13} M_{\odot}$ for the classical case
$\Lambda = 0 $, and  $1.25\cdot 10^{13} M_{\odot}$ for $\Omega_{\lambda} = 0.73$.
Because the mean radius $R_o$ for the groups exceeds  5  times their mean
virial radius, the agreement of independent internal and external mass
estimates may be interpreted as the absence of a dark matter outside the
$R_{vir}$. This unexpected result should be proven by new observational
data.

{\bf Acknowledgements} This work was supported by DFG--RFBR grant 06--02--04017
and RFBR grant 07-02-00005.

 {}
\end{document}